\documentclass{appolb}
\usepackage{graphicx}
\usepackage{amsmath, amssymb}
\usepackage{physics}
\usepackage{bbold}
\usepackage{xcolor}
\usepackage{authblk}
\usepackage{mathtools}
\usepackage{subcaption}
\usepackage{cite}
\usepackage{soul}
\usepackage{array} % for defining new column types

\newcommand{\rev}[1]{{\color{black} #1}}  %colors revisions as red, comment out for clean copy
\newcolumntype{P}[1]{>{\centering\arraybackslash}p{#1}}

\begin{document}
\title{
Universal Frequency Correlations and Recurrence Statistics of Complex Impedance Matrices
}

\author[1]{Nadav Shaibe \thanks{Corresponding author: nshaibe@umd.edu}}
\author[1]{Jared Erb}
\author[1,2]{Thomas M. Antonsen}
\author[1,2]{Steven M. Anlage}

\affil[1]{\footnotesize Department of Physics, University of Maryland, College Park, Maryland 20742-4111, USA}
\affil[2]{\footnotesize Department of Electrical and Computer Engineering, University of Maryland, College Park, Maryland 20742-3285, USA}

\maketitle
\begin{abstract}
Linear electromagnetic wave scattering systems can be characterized by an impedance matrix that relates the voltages and currents at the ports of the system. When the system size becomes greater than the wavelength of the fields involved, the impedance matrix becomes a complicated function of the details of the system, in which case a statistical model, such as the Random Coupling Model (RCM) becomes useful. The statistics of the  elements of the RCM impedance matrix depend on the excitation frequency, the spectral density of the modes of the enclosed system volume, the average loss factor ($Q^{-1}$) of the system, and the properties of the coupling ports as given by their radiation impedances. In this paper, properties of the elements of impedance matrices are explored numerically and experimentally. These include the two point frequency correlation functions for the complex impedance of elements and the expected difference in frequencies between which impedance values are approximately repeated. Universal scaling arguments are then given for these quantities, hence these results are generic for all sufficiently complicated scattering systems, including acoustic and optical systems. The experimental data presented in this paper come from microwave graphs, billiards, and three-dimensional cavities with embedded tunable perturbers such as metasurfaces. The data is found to be in generally good agreement with the predictions for the two point frequency correlations and the frequency interval for successive repetitions of impedance matrix elements values.
\end{abstract}
  
\section{Introduction}

We consider the propagation of waves through complicated scattering systems with a finite number of asymptotic scattering channels coupled to the outside world. Generically the systems could be three-dimensional spaces, two-dimensional billiards, or one-dimensional graphs, and the waves could be electromagnetic, acoustic, quantum, etc. When the dimensions of the systems are much larger than the wavelength of waves propagating within, the scattering properties become extremely sensitive to cavity details, such as the shape of the boundaries and the presence of internal scatterers. Because of this sensitivity and the resulting high variability of wave properties, these systems are referred to as being ``wave chaotic".  As a result, deterministic calculations for the properties of a single realization of such a system are difficult to obtain, leading to the use of statistical approaches to calculating ensemble properties.

Random Matrix Theory (RMT) based models have emerged as especially successful at predicting certain universal properties of wave chaotic systems \cite{Wigner1947,Lyon1975,Couchman1992,Beenakker1997,Guhr1998,Guhr1989,Alhassid2000,Mendez2003,Fyodorov2004,Kuhl2005,Weidenmuller2009,Mitchell2010,Langley2010,Lawniczak2011,Maksimov2011,Kuhl2016,Li2017,Grabsch2018,Hougne2020}.  One such model, the Random Coupling Model (RCM), was developed to incorporate nonuniversal physical features that practical systems invariably have, such as coupling details of ports or antennas \cite{Zheng20051Port,Zheng20052Port} and short orbits (also known as direct processes) \cite{Hart2009,Yeh2010,Lee2013}, with an RMT description of the wave propagation within the enclosure. The RCM models the system including its ports in terms of an impedance or admittance matrix that linearly relates the voltages and currents appearing at the system's ports.  From these matrices, combined with knowledge of the external transmission systems feeding the ports, it is possible to evaluate the system scattering matrix, which is often of interest.  

The RCM has been applied to understanding the statistical electromagnetic properties of complex enclosed systems for decades \cite{Hemmady2005,GilGil2016,Ma2023b}, and has also been empirically verified to describe acoustic scattering \cite{Auregan16}. Predictions of the RCM have been compared to experimental scattering and impedance matrices \cite{Hemmady2005,Hemmady2005b,Yeh2010,Yeh2010b, Lawniczak2010,Lawniczak2019}, admittance matrices \cite{Hemmady2006c}, conductance fluctuations \cite{Hemmady2006}, fading statistics \cite{Yeh2012}, complex time delay statistics \cite{Lei2021Stats,Shaibe2025_CTD}, and scattering singularity statistics \cite{Shaibe2025_TOP}.

One of the applications of the RCM is identifying the underlying universal fluctuations of impedance from experimental ensemble data, which is done by normalizing the measured cavity impedance matrix by the \textit{radiation impedance} specific to the ports or antennas used in the measurement. \rev{The radiation impedance of an antenna characterizes how well it couples to electromagnetic fields in an environment where enclosing boundaries have been removed. This provides specific baseline characterization of the coupling at a port using that antenna. One way to measure the radiation impedance is to put the antenna in an anechoic chamber, equivalent to letting the antenna radiate out to infinity with no reflections, and measuring the resulting scattering parameters. The radiation impedances of the ports are then used to normalize the actual impedance matrix, yielding a dimensionless impedance matrix $Z$ that is independent of many details of the system, allowing for comparison with universal RMT predictions}. In this paper, we examine the RCM normalized impedance matrix obtained from experimental data, and identify characteristics of the two point frequency correlation functions of the real (resistance $R$) and imaginary (reactance $X$) components of the impedance matrix elements.

\begin{figure}[hbt]
\includegraphics[width=\textwidth]{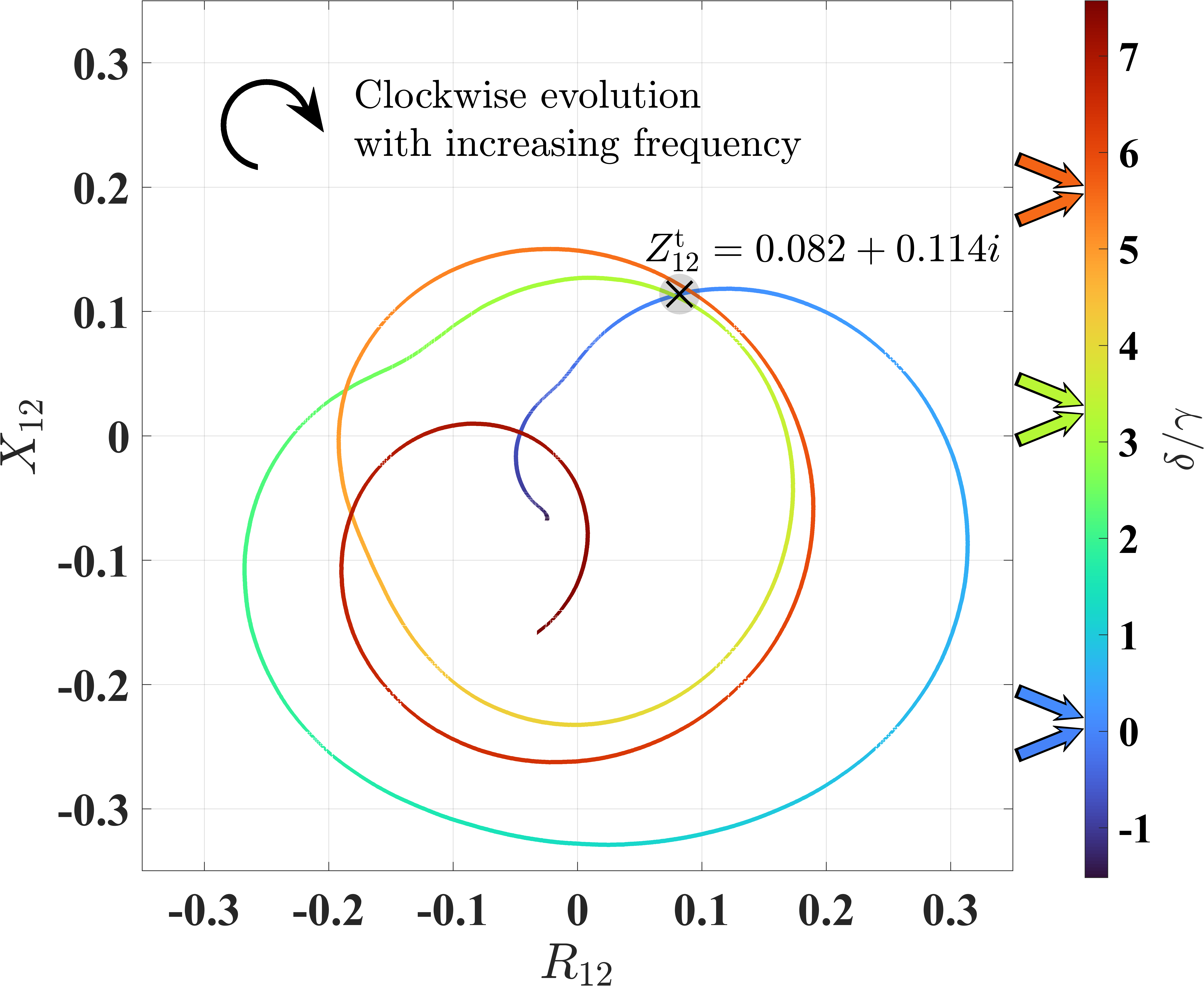}
\caption{Clockwise path traced by normalized $Z_{12}$ in the complex impedance plane over a small frequency band. The color scale corresponds to the frequency difference $\delta=f_a-f_b$ where $f_b$ is the first frequency at which $Z_{12}$ is within the gray circle of radius $\epsilon=0.015$ centered on $Z_{12}^\text{t}=0.082+0.114i$. $\delta$ is normalized by a characteristic scale corresponding to the typical $Q$-widths of the modes $\gamma=2\alpha\Delta=0.66~\text{MHz}$, where $\alpha$ is the RCM loss parameter and $\Delta$ is the mean mode spacing. The arrows on the color bar mark the frequencies at which $Z_{12}$ enters and leaves the circle. The impedance $Z_{12}$ plotted in this figure is experimentally measured from a three dimensional cavity with absorption $\alpha=5.5$.}
\label{Fig_ZPlane}
\end{figure}

We then apply the knowledge of the correlation functions to address the question: If a port of the system presents a particular value of the complex impedance $Z^\text{t}$ at a frequency $f$,  what is the average frequency interval $\Lambda$ before the same value of impedance $Z^\text{t}$, within some specified tolerance, is seen again? An example is shown in Fig.~\ref{Fig_ZPlane} where we show the path in the complex plane traced by an off-diagonal element of the RCM normalized impedance matrix, $Z_{12}$,  over a short frequency range. The arbitrarily chosen target impedance value is $Z_{12}^\text{t}=0.082+0.114i$ marked by the black cross, and the small gray circle around it represents a tolerance threshold of $\epsilon=0.015$. The color bar represents the frequency difference $\delta=f_a-f_b$ (where $f_b$ is the first frequency at which $Z_{12}$ is within the gray circle) normalized by a characteristic scale corresponding to the typical $Q$-widths of the modes $\gamma=2\alpha\Delta=0.66~\text{MHz}$. \rev{H}ere $\alpha$ is the dimensionless RCM loss parameter \rev{which represents the degree of overlap of the modes ($Q$-width normalized to the mean mode spacing), while} $\Delta$ is the mean frequency spacing of the modes. 

Empirically we find that the expected frequency interval $\Lambda$ between repetitions of a target impedance depends on the mean mode spacing $\Delta$ of the system, the degree of absorption $\alpha$ of the system, the value of the desired target impedance $Z^\text{t}$, and the tolerance $\epsilon$ on the target value. We claim the scaling of $\Lambda$ to be significant for several reasons. $\Lambda$ is a new statistical observable for chaotic scattering \rev{that plays a similar role to level-spacing statistics for eigenfrequencies, encoding how frequency correlations determine the recurrence of observable quantities. $\Lambda$ is} directly analogous to \rev{the Poincar\'e} recurrence time in \rev{a non-autonomous, 2 degree of freedom, continuous time} dynamical chaotic system, \rev{and also parallels the mean first-passage time in stochastic processes}. Our scaling law provides a predictive tool for how often a desired impedance condition occurs naturally in frequency for a given cavity or resonator\rev{, which} has practical relevance for system engineering. \rev{If, for example, one wants to use impedance as a sensor, such as of gas concentration or cavity defect formation, by comparing a measured impedance value with a target $Z^\text{t}$, then $\Lambda$ would be an important quantity to consider. A large recurrence interval would mean the impedance spectrum is smooth with only a few and distant features, making impedance not useful for such applications due to a lack of sensitivity to perturbations. But with too small of a $\Lambda$, one may encounter issues because the impedance value naturally repeats with even minor perturbations.} This is also the first step towards a more difficult question, which would be looking for repetitions not of individual impedance elements, but of special conditions on the entire $Z$ matrix, such as impedance matrix Exceptional Points. \rev{Exceptional Point sensors have been proposed which take advantage of the sublinear splitting of the eigenvalues upon perturbation, making them the subject of considerable interest for a variety of sensing applications \cite{Wiersig2014,Zhicheng2019,Chen2025}.} Impedance matrix Exceptional Points have previously been shown to be identical to scattering matrix Exceptional Points \cite{Erb2025}, so the impedance matrix may serve as a useful platform to study these and other wave scattering singularities.

In this manuscript, we first present the theoretical predictions for the correlation functions and repetition interval $\Lambda$ in Sec.~\ref{SEC_THEORY}. The experimental systems used are introduced in Sec.~\ref{SEC_EXP}, alongside an explanation of how we apply the Random Coupling Model to the raw data and the statistical methods used in the analysis. The results are shown and interpreted in Sec.~\ref{SEC_DISCUSSION}, and we conclude our work and discuss future research in Sec.~\ref{SEC_CONCLUSION}. \rev{In Table~\ref{TABLE_PARA} of Appendix A we list the relevant parameters that are utilized in this work.}

\section{Theory}\label{SEC_THEORY}

In this section, we define the quantities important to this paper, and through simple arguments derive theoretical predictions to compare with our empirical results. The frequency response of an over moded, multiport wave enclosure characterized by an impedance matrix is well captured by the RCM.  As frequency varies, so do the values of the elements of the impedance matrix.  These variations can be characterized in multiple ways.  We have considered two such characterizations. First is the two point frequency correlation functions for the real and imaginary fluctuating components of the impedance matrix elements $\langle \tilde{R}(f_a)\tilde{R}(f_b)\rangle_C$, $\langle \tilde{X}(f_a)\tilde{X}(f_b)\rangle_C$, and $\langle \tilde{R}(f_a)\tilde{X}(f_b)\rangle_C$, where angled brackets with subscript $C$ means  the two point correlation averaged over realizations). Notice that the tilde notation, such as $\tilde{R}$, represents the fluctuations about the mean value of value, in this case of $R$. It so happens that for any off-diagonal element of the RCM normalized impedance matrix, the mean of $Z_{pp'}$ is $0+0i$ so $R_{pp'}=\tilde{R}_{pp'}$ and $X_{pp'}=\tilde{X}_{pp'}$ ($p\neq p'$). However, for the diagonal elements, the mean of $Z_{pp}$ is $1+0i$, meaning $X_{pp}=\tilde{X}_{pp}$ but $R_{pp}\neq\tilde{R}_{pp}$. We therefore suppress the tilde notation for $X_{pp}$, $R_{pp'}$, and $X_{pp'}$, and keep it only for $\tilde{R}_{pp}$. The reason for looking at the correlation functions of the fluctuations is to keep the scales of the two point frequency correlation functions consistent.

General forms of two point frequency correlation functions for the fluctuations of the impedance and scattering matrices were first proposed by Fyodorov, Savin, and Sommers \cite{Fyodorov2005,Savin2006}. Various scattering matrix correlation functions have been studied since, often as a way to probe what is known as the elastic enhancement factor \cite{Hemmady2006c,Dietz2009,Dietz2010,Lawniczak2010,Lawniczak2011,Lawniczak2012,Yeh2013,Lawniczak2015,Sokolov2015,Ericson2016,Ramos2016,Bialous2020,Bereczuk2021,Davy2021,Bialous2023,Lawniczak2023}. In this paper, we consider the impedance two point autocorrelation functions, which have not been experimentally investigated before, to our knowledge. 

Under the limiting assumptions of ideal coupling (which we achieve through the RCM normalization process), reciprocity (Dyson symmetry class $\beta=1$), and large loss ($\alpha\gg1$, \rev{meaning the modes are strongly overlapping}), we arrive at the reduced expressions for the impedance two point frequency correlation functions for the elements of the RCM normalized impedance matrix with condensed notation:
\begin{equation}
    \langle \tilde{R}_{pp}(f_a)\tilde{R}_{pp}(f_b)\rangle_C = \langle X_{pp}(f_a)X_{pp}(f_b)\rangle_C =2(R^{\text{RMS}})^2\frac{1}{(\delta/\gamma)^2+1},\label{EQN_CorrR}
\end{equation}
\begin{equation}
       \langle \tilde{R}_{pp}(f_a)X_{pp}(f_b)\rangle_C = -2(R^{\text{RMS}})^2\frac{\delta/\gamma}{(\delta/\gamma)^2+1},\label{EQN_CorrRCross}
\end{equation}
\begin{equation}
    \langle R_{pp'}(f_a)R_{pp'}(f_b)\rangle_C = \langle X_{pp'}(f_a)X_{pp'}(f_b)\rangle_C=(R^{\text{RMS}})^2\frac{1}{(\delta/\gamma)^2+1},\label{EQN_CorrT}
\end{equation}
\begin{equation}
    \langle R_{pp'}(f_a)X_{pp'}(f_b)\rangle_C = -(R^{\text{RMS}})^2\frac{\delta/\gamma}{(\delta/\gamma)^2+1}.\label{EQN_CorrTCross}
\end{equation}
In these equations, $\delta=|f_a-f_b|$ is the difference in frequency between the two points, $\gamma=2\alpha\Delta$ is the characteristic frequency interval specific to the system, and $R^{\text{RMS}}= \sqrt{\langle (R_{pp'})^2\rangle}$ is defined as the standard deviation of the fluctuation of the real component of an off-diagonal element of the RCM normalized impedance matrix. Note that in the case of large $\alpha$, $R^\text{RMS}$ is known to depend only on $\alpha$ by the formula $R^\text{RMS}=(\pi\alpha)^{-1/2}$ \cite{Hemmady2005,Zheng20051Port}.

\begin{figure}[htb]
\centerline{
\includegraphics[width=13cm]{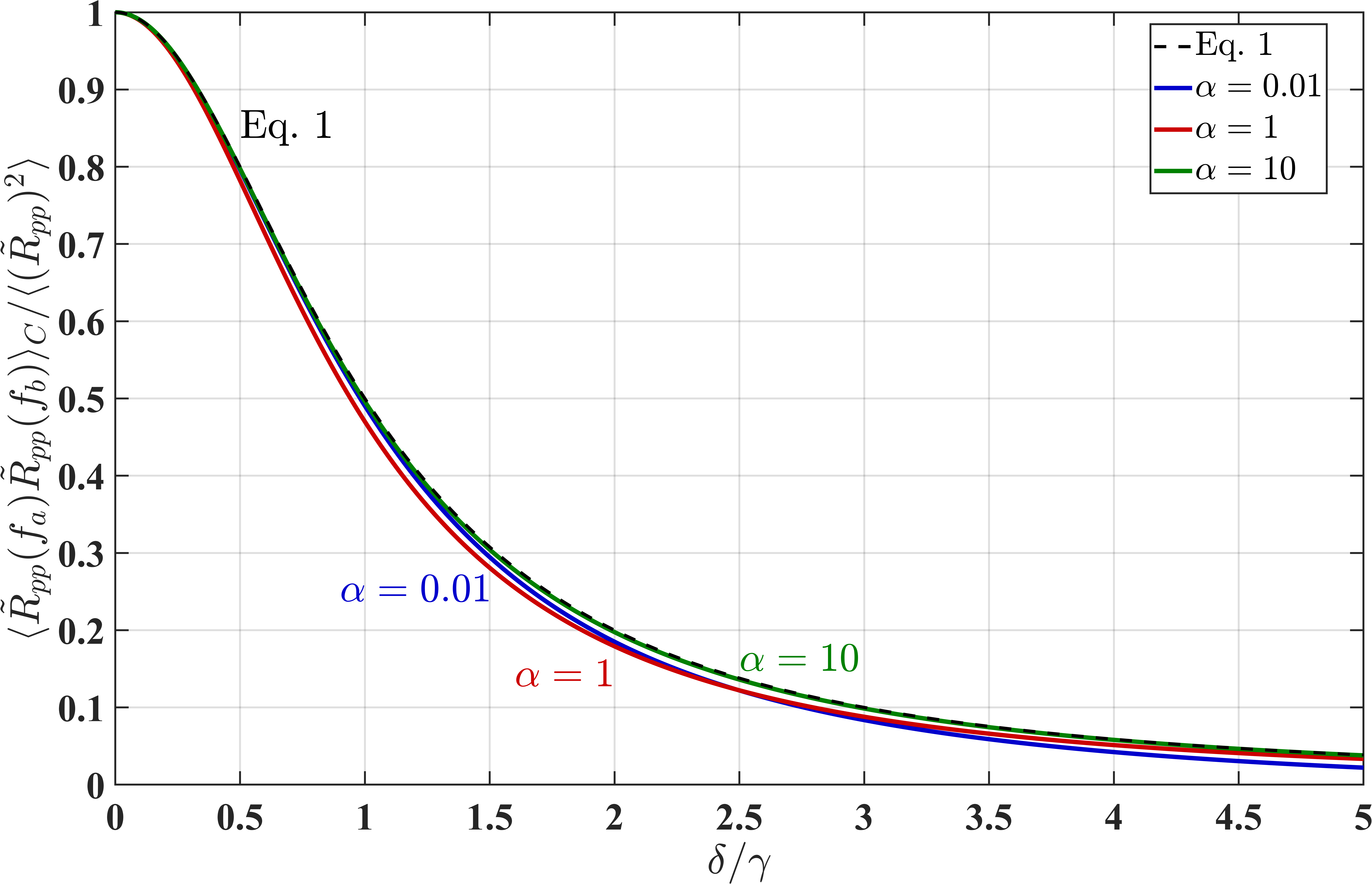}}
\caption{Two point frequency correlation function of the real part of a diagonal impedance element $\langle\tilde{R}(f_a)\tilde{R}(f_b)\rangle_C$ normalized by the zero separation correlation $\langle (\tilde{R}_{pp})^2 \rangle$ as a function of $\delta/\gamma=\frac{|f_a-f_b|}{2\alpha\Delta}$. Comparing prediction of Eq.~\ref{EQN_CorrR} illustrated by the dashed black curve, and expression derived by Fyodorov, Savin, and Sommers in Ref.~\cite{Fyodorov2005}, represented by the colored curves corresponding to different choices of loss parameter $\alpha$. The green curve which has $\alpha=10$ is indistinguishable from the dashed black curve.}
\label{Fig_Fyo}
\end{figure}

The correlation functions for the diagonal elements come out to be exactly twice those of the off-diagonal elements, and there is no difference between the real and imaginary components. The autocorrelation of any component is peaked at zero frequency separation, while the cross correlation of a real component with an imaginary component is zero at zero frequency separation. The negative sign in Eqs.~\ref{EQN_CorrRCross} and \ref{EQN_CorrTCross} is a matter of convention for the phase evolution of $Z(f)$ \cite{Lei2021Stats}. Because microwave network analyzers utilize a convention in which the phase decreases with increasing frequency, corresponding to the clockwise motion of $Z(f)$ in the $R+iX$ plane seen in Fig.~\ref{Fig_ZPlane}, we include the negative sign.

In Fig.~\ref{Fig_Fyo} we compare Eq.~\ref{EQN_CorrR} to the \rev{correlation function of the real part of the diagonal impedance using} Eqs.~[10a-b] of Ref.~\cite{Fyodorov2005} with $\beta=1$ (reciprocity), all $\kappa=1$, (ideal coupling), and three different $\alpha$ values. We see that for large $\alpha$, our equation and the expression derived by Fyodorov, Savin, and Sommers for the two point frequency correlation function of the real part of the diagonal impedance elements are equivalent. \rev{Notice that the correlation function does not change dramatically with $\alpha$, so while our expressions are not perfect they can function as fair approximations even at medium to low loss and weakly overlapping modes. This is true not only in the case of $\langle \tilde{R}_{pp}(f_a)\tilde{R}_{pp}(f_b)\rangle_C$, but also for the correlation function of the imaginary part of the diagonal impedance $\langle X_{pp}(f_a)X_{pp}(f_b)\rangle_C$. When it comes to the off-diagonal impedance, the correlation functions are less sensitive to $\alpha$, so Eqs.~\ref{EQN_CorrT}-\ref{EQN_CorrTCross} should be even more successful approximations than Eqs.~\ref{EQN_CorrR}-\ref{EQN_CorrRCross} at low loss.}

\rev{The absence of substantial change in the two point frequency correlation functions with $\alpha$ tells us something profound and unexpected. At small $\alpha$, the modes have a narrow $Q$-width relative to the mode spacing. The correlation function over a small frequency band should then look like the correlation function of an isolated mode uninfluenced by any neighbors, which would be a Lorentzian with a scale of the mode $Q$-width $\gamma$. In Fig.~\ref{Fig_Fyo}, we see that the correlation width is nearly the same for overlapping as well as for non-overlapping modes.}

The second statistical characterization we consider, motivated by the path traced by $Z(f)$ in the complex plane as frequency is varied, is how often an element of the impedance matrix comes within a small distance $\epsilon$ of a target value $Z^\text{t}$. If the cavity losses are large, such that the $Q$-widths of modes \rev{$\gamma$} are larger than the frequency spacing between modes \rev{$\Delta$}, the two point frequency impedance correlation functions decay in a frequency interval $\gamma$. Thus one can estimate that over a frequency range of $\gamma$, the impedance value has traced a path of length $R^\text{RMS}$. We now imagine that the path in the $R+iX$ plane, rather than being a one-dimensional line, is broadened to a width $\epsilon$. After $N$ frequency intervals of length $\gamma$, the broadened path will cover an area $N\epsilon R^\text{RMS}$. That path is confined approximately to be within a circle of radius $R^\text{RMS}$. Equating the area covered by the path to the area $\pi (R^\text{RMS})^2$ within a circle of radius $R^\text{RMS}$, we determine the expected number $N$ of frequency intervals of length $\gamma$, needed for the impedance value to come within $\epsilon$ of a target value $Z^\text{t}$ within the circle, which is $N=\frac{\pi R^\text{RMS}}{\epsilon}$. This gives a simple estimate for the typical frequency interval needed for the impedance to come within $\epsilon$ of a target value as $\Lambda = \frac{\pi\gamma R^\text{RMS}}{\epsilon}$.

However, the target impedance $Z^\text{t}$ could be further than $R^\text{RMS}$ from the starting impedance, and the further the distance the greater the increase in the expected frequency interval. We estimate the increase to be proportional to the ratio of the peak of the probability density function (PDF) of impedance $\mathcal{P}(Z^\text{peak})$ to the value of the PDF at the target impedance value $\mathcal{P}(Z^\text{t})$. The expected frequency interval is then given by:
\begin{equation}
    \Lambda(Z^\text{t}) = \pi\frac{\gamma R^\text{RMS}}{\epsilon}\frac{\mathcal{P}(Z^\text{peak})}{\mathcal{P}(Z^\text{t})}. \label{EQN_FreqInterval}
\end{equation}

This quantity is defined to be the average frequency interval between two consecutive visits of the system impedance to be within distance $\epsilon$ of a given target impedance $Z^\text{t}$. This equation can be rewritten slightly to be more explicit by using the high loss approximation of $R^\text{RMS}$:
\begin{equation}
    \Lambda(Z^\text{t}) = 2\left(\pi\alpha\right)^{1/2} \frac{\Delta}{\epsilon}\frac{\mathcal{P}(Z^\text{peak})}{\mathcal{P}(Z^\text{t})}. \label{EQN_FreqInterval_SIMP}
\end{equation}
We conclude that the frequency interval $\Lambda$ scales proportionally with the mean mode spacing $\Delta$ and inversely with the tolerance $\epsilon$. The dependence on loss is more complicated, however, as there is a hidden and non-trivial $\alpha$ dependence in the distribution function $\mathcal{P}(Z)$. Our theory results apply in the over moded limit, where the spacing between modes \rev{$\Delta$} is much less than the typical $Q$-width of the modes \rev{$\gamma$}. This corresponds to the RCM loss parameter $\alpha\gg1$. In this case multiple modes contribute to the response at any frequency, the details of the distribution of mode spacings become unimportant, and the values of the impedance matrix elements become Gaussian random variables. We can approximate the probability ratio in this limit ($\alpha\gg1$) to take the form
\begin{equation}
    \frac{\mathcal{P}(Z^\text{peak})}{\mathcal{P}(Z^\text{t})}
    \approx
    \mathcal{P}(Z^\text{peak}) 
    \begin{cases}
    \exp\left[
        \frac{(X^\text{t})^2 + (R^\text{t}-1)^2}{(R^\text{RMS})^2}
    \right], & \text{Diagonal} \\
      \,
    \exp\left[
        \frac{(X^\text{t})^2 + (R^\text{t})^2}{(R^\text{RMS})^2}
    \right], & \text{Off-Diagonal}
    \end{cases}
    \label{EQN_PDF}
\end{equation}
depending on \rev{whether it is a diagonal or off-diagonal element of the impedance matrix that is being considered}.

\section{Experimental Setup and Data Analysis}\label{SEC_EXP}

\begin{figure}[hbt]
\includegraphics[width=\textwidth]{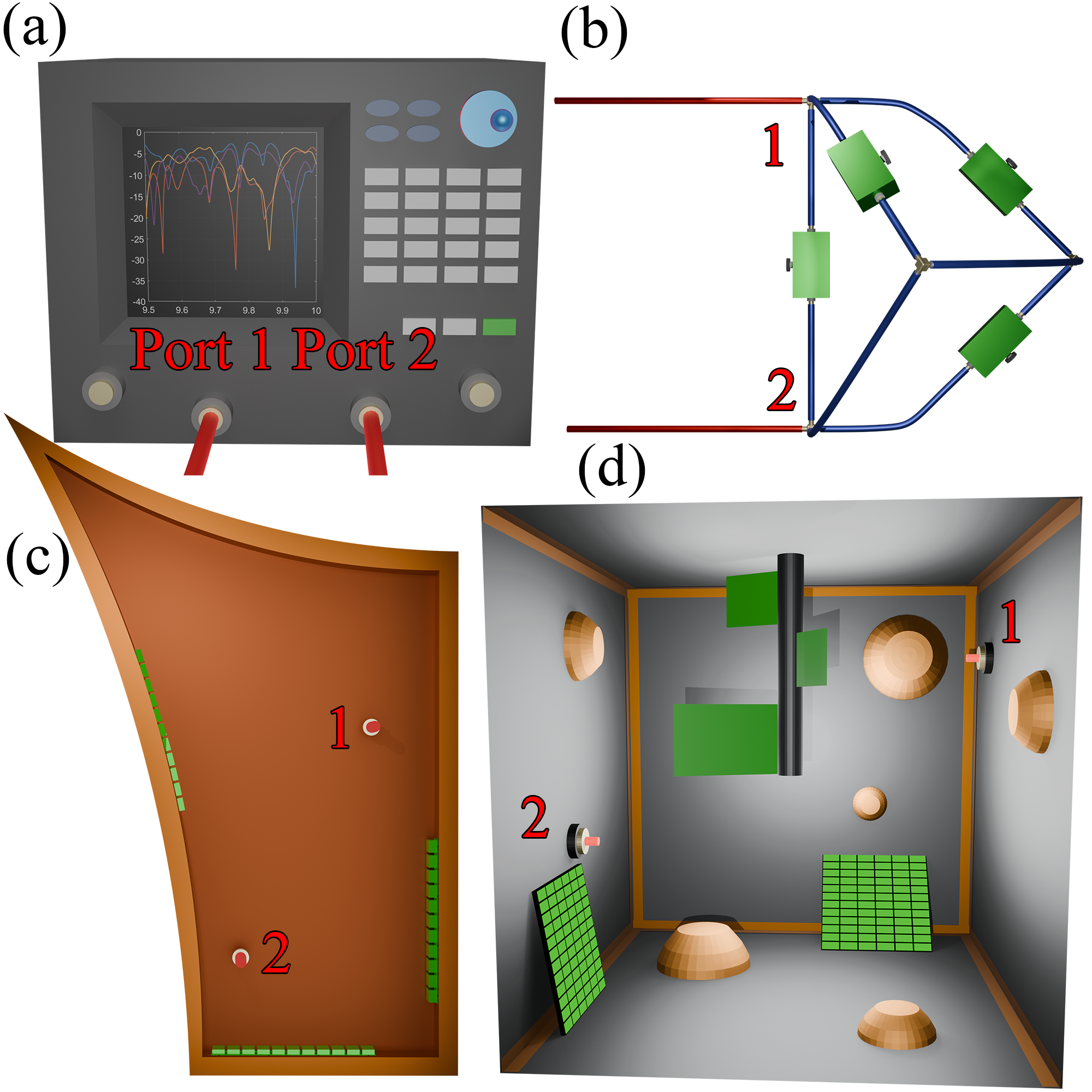}
\caption{(a) Vector network analyzer with four total ports. Two of the ports are connected through the red cables to the following experimental systems: (b) a tetrahedral microwave graph ($\mathcal{D}=1$), (c) a ray-chaotic quarter bowtie billiard ($\mathcal{D}=2$), and (d) a three dimensional cavity with various symmetry breaking elements ($\mathcal{D}=3$). Embedded tunable perturbers used for ensemble creation are marked in green.}
\label{Fig_Schematic}
\end{figure}

In this section we introduce the experimental platforms and methods used to verify the theoretical predictions in Sec.~\ref{SEC_THEORY}. Our empirical data comes from three kinds of microwave systems which we characterize by the number of dimensions of wave propagation $\mathcal{D}$. Schematics of the systems are provided in Fig.~\ref{Fig_Schematic}: (b) a tetrahedral graph ($\mathcal{D}=1$) \cite{Kottos1997,Kottos2000,Hul2004,Hul2005}, (c) a chaotic quarter bowtie billiard ($\mathcal{D}=2$) \cite{Doron1990,Sridhar1991,Graf1992,Alt1998,Stockmann1999,Stockmann1999,Kumar2013,Dietz2015}, and (d) a three dimensional cavity with various geometrical symmetry breaking elements ($\mathcal{D}=3$) \cite{Alt1997,Lawniczak2009,Selemani2015}. The microwave scattering within these cavities is considered complex when the systems are excited with waves whose wavelengths are small compared to the system size. The interference effects for waves following different ray trajectories are then extremely sensitive to details of the system configuration. We use embedded tunable perturbers to electronically change the system configuration \textit{in situ} in between measurements. The tunable perturbers are illustrated in green in Fig.~\ref{Fig_Schematic}. In a graph, we use voltage controlled mechanical phase shifters \cite{Lawniczak2023} which act as variable-length cables, allowing us to change the interference conditions at the nodes. In the billiard and three dimensional cavity we use globally-biased, voltage controlled varactor-loaded metasurfaces \cite{Sleasman2023,Erb2024} which change the amplitude and phase of reflected waves. In the three dimensional cavity we also have a mechanical mode stirrer attached to the ceiling which can be rotated, substantially altering the wave propagation paths. In reverberant scattering environments, most waves will interact with these tunable scattering elements multiple times, allowing for strong control over the wave scattering properties of each system.

The systems are connected to a Keysight PNA-X microwave network analyzer through asymptotic channels marked in red in Fig.~\ref{Fig_Schematic}. The calibrated network analyzer measures the scattering matrix $S^{\text{raw}}$ of the system, and having at least two ports allows for both reflection and transmission measurements. To investigate statistics, an ensemble is made by measuring the same system hundreds of times with different configurations of the embedded perturbers. The $S^{\text{raw}}$ matrices from the network analyzer are then converted to $Z^{\text{raw}}$ matrices by
\begin{equation}
    Z^{\text{raw}}=Z_0^{1/2}\frac{I_{M\times M}+S^{\text{raw}}}{I_{M\times M}-S^{\text{raw}}}Z_0^{1/2} \label{EQN_S2Z}
\end{equation}
where $Z_0$ is a diagonal matrix of the characteristic impedances of the scattering channels ($50~\Omega$, in our case), and $M$ is the number of scattering channels. We then calculate the universal fluctuating impedance $Z$ \rev{with system-specific features normalized} via the formula \cite{Hemmady2005,Zheng20051Port,Zheng20052Port}
\begin{equation}
    Z=(\text{Re}[\langle Z^{\text{raw}}\rangle])^{-1/2}(Z^{\text{raw}}-i\text{Im}[\langle Z^{\text{raw}}\rangle])(\text{Re}[\langle Z^{\text{raw}}\rangle])^{-1/2}, \label{EQN_NORMZ}
\end{equation}
where $\langle Z^{\text{raw}}\rangle$ is a matrix whose elements are the ensemble averages of the elements of $Z^{\text{raw}}$. \rev{In an ensemble with a high degree of statistical independence between the realizations, we can expect $\langle Z^{\text{raw}}\rangle$ to approach the radiation impedance in the limit of the number of realizations going to infinity, though for practical purposes a few hundred realizations is almost always sufficient. The advantage of using the ensemble average over the radiation impedance in the normalization process is that $\langle Z^{\text{raw}}\rangle$ also contains information about the short orbits of the cavity, which are another source of non-universal effects \cite{Yeh2010,Yeh2010b}.} An estimate for the cavity loss $\alpha$ of the system can then be obtained by simultaneously matching the $2M^2$ PDFs of real and imaginary components of the RCM normalized impedance matrix to those predicted by Random Matrix Theory \cite{Gradoni2014,Fu2017}
\begin{equation}
    Z^{\text{RCM}} = -\frac{i}{\pi}\sum_{n}\frac{W_nW_n^T}{\zeta_n^{\text{RMT}}+i\alpha}.\label{EQN_ZRCM}
\end{equation}
 $\zeta_n^{\text{RMT}}$ is the $n^\text{th}$ eigenvalue of an $N\times N$ random matrix used as the Hamiltonian of the closed system, and the vector $W_n$ is the $n^\text{th}$ column of the $M\times N$ matrix which describes the coupling between the $N$ closed cavity modes to the $M$ channels. 
 
 Experimentally, we made ensembles with different values of $\alpha$ by measuring in different frequency bands, changing the overall size of the systems, and distributing attenuators on the bonds of a graph \cite{Hul2005,Lawniczak2023,Lawniczak2019}. More details about the experimental systems used in this paper are discussed in Refs.~\cite{Erb2025,Shaibe2025_CTD,Shaibe2025_TOP}, and an example of determining an experimental ensemble $\alpha$ value through the RCM normalization process is provided in Appendix B.

The equations presented in Sec.~\ref{SEC_THEORY} were derived with the assumption of high loss ($\alpha\gg1$). Because cavity loss scales with cavity size, the dimensionality parameter $\mathcal{D}$ naturally separates our ensembles into different regimes. The tetrahedral graph ($\mathcal{D}=1$) has the lowest loss with typical $\alpha$ values between $0.2$ and $2$, depending on overall length, frequency band, and inclusion of attenuators on the bonds. The quarter bowtie billiard ($\mathcal{D}=2$) has moderate loss with $\alpha$ being around $3$ in the frequency bands where the metasurfaces are most effective. The three dimensional cavity ($\mathcal{D}=3$) is the only system for which we can get $\alpha>4$ and maintain good tunability of our perturbers for high quality ensembles. We will still present data from the $\mathcal{D}=1$ and $\mathcal{D}=2$ systems to examine how our theory extends beyond the absorption limit we expect it to be valid for.

Because all diagonal elements $Z_{11},Z_{22},\dots,Z_{MM}$ are statistically identical, only one of these quantities needs to be considered, and the same is true for the off-diagonal elements. This means $M=2$ channels is sufficient to show all the unique statistics in this paper. Note that this is only because we are using the RCM normalized impedance, and is not necessarily true for a raw impedance matrix \rev{which would have system-specific features that could be different at each port}.

\begin{figure}[htb]
\centerline{
\includegraphics[width=13cm]{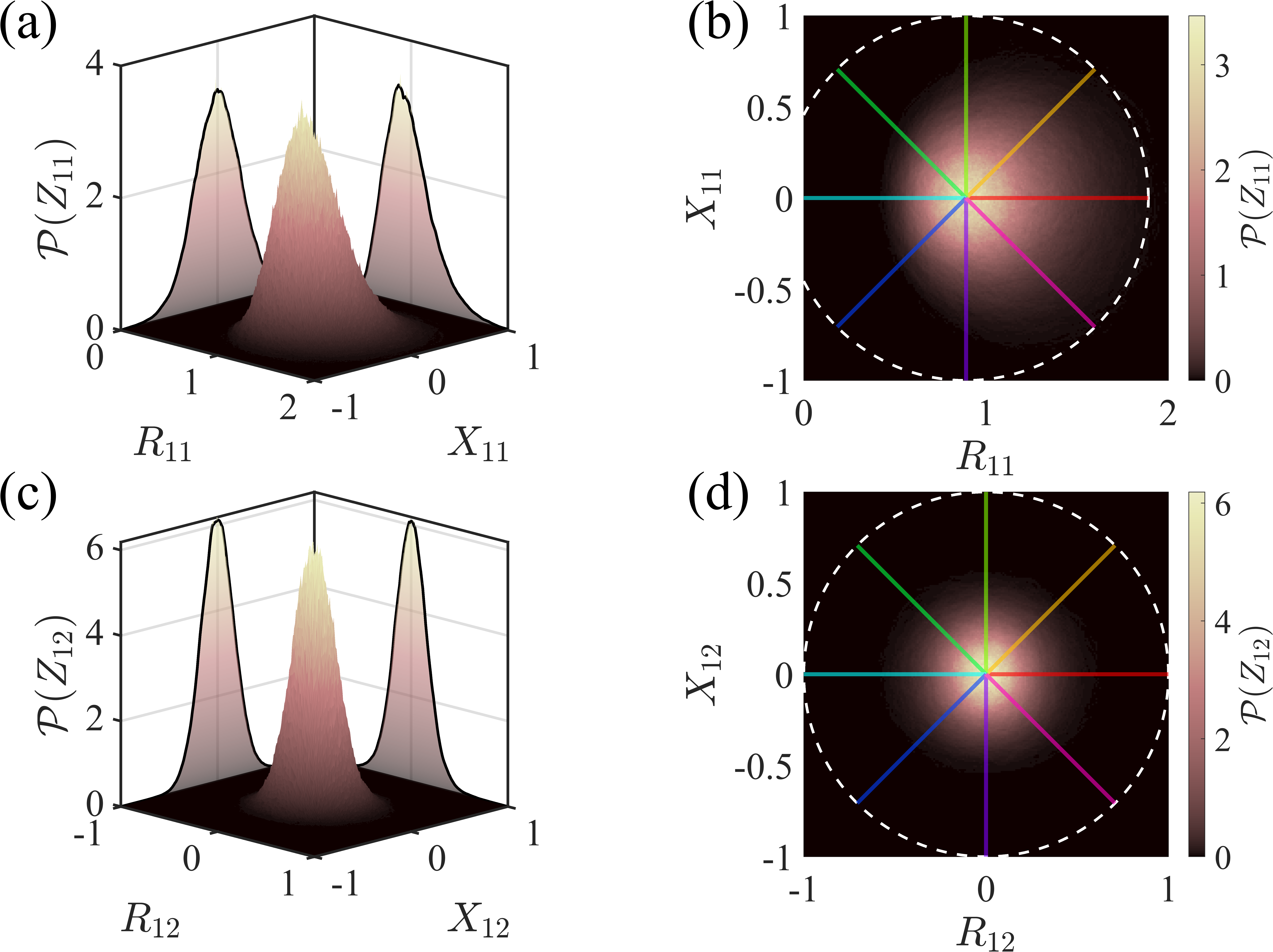}}
\caption{Bivariate PDFs of complex impedance $Z$ at port 1 of a $\mathcal{D}=3$ cavity with absorption $\alpha=5.5$. (a) PDF of diagonal impedance $Z_{11}$, projections show the PDF is symmetric along $X_{11}$ but asymmetric along $R_{11}$. (b) Top down view of $\mathcal{P}(Z_{11})$ where the colored lines correspond to trajectories of target impedance $Z^\text{t}_{11}$ away from $Z^\text{peak}_{11}$ along different angles $\theta=\text{Arg}(Z^\text{t}_{11}-Z^\text{peak}_{11})$. (c) PDF of off-diagonal impedance $Z_{12}$, projections show the PDF is symmetric along both $X_{12}$ and $R_{12}$. (d) Top down view of $\mathcal{P}(Z_{12})$ where the colored lines correspond to trajectories of target impedance $Z^\text{t}_{12}$ away from $Z^\text{peak}_{12}$ along different angles $\theta=\text{Arg}(Z^\text{t}_{12}-Z^\text{peak}_{12})$.}
\label{Fig_PDF}
\end{figure}

We use the RCM normalized impedance matrix diagonal ($Z_{11}$) and off-diagonal ($Z_{12}$) elements to calculate the two point frequency correlation functions by calculating the correlation function for each individual realization and then averaging those correlation functions (Eqs.~\ref{EQN_CorrR}-\ref{EQN_CorrTCross}) over the ensemble for each frequency difference $\delta$. To calculate the frequency interval $\Lambda$ between repeated values of $Z^\text{t}$, we take the total bandwidth of the ensemble, defined as the frequency band of a single realization times the number of measurements in the ensemble, and divide by the total number of unique times $|Z-Z^\text{t}|\leq\epsilon$ over the entire ensemble. If $Z$ is within $\epsilon$ of $Z^\text{t}$ for multiple consecutive frequency points, we only count that as one event. We do this for $100$ values of $\epsilon$ that are evenly spaced on a logarithmic scale between $\epsilon=10^{-5}$ and $\epsilon=10^{-1}$, and for $241$ values of $Z^\text{t}$, for both $Z_{11}$ and $Z_{12}$. The values of $Z^\text{t}$ used in this paper are $Z^\text{peak}$ (the value of $Z$ for which $\mathcal{P}(Z)$ is maximized) and thirty points each along eight trajectories that have angle $\theta=[0,\pi/4,\pi/2,3\pi/4,\pi,5\pi/4,3\pi/2,7\pi/4]$ in the $R+iX$ plane.

In Fig.~\ref{Fig_PDF}(a) and (c), we show the PDFs of $Z_{11}$ and $Z_{12}$ from a three dimensional microwave cavity with absorption parameter $\alpha=5.5$. As can be seen by the projections on the far surfaces of the panels, $\mathcal{P}(Z_{11})$ is not symmetric in the $R_{11}$ direction. In a passive lossy system, it is required that $R_{pp}\geq0$ but simultaneously $\langle R_{pp}\rangle=1$, hence the peak of the PDF is shifted to smaller values and has a long tail. The smaller the $\alpha$, the closer the PDF peak shifts towards $R_{pp}=0$. In contrast, $\mathcal{P}(Z_{12})$ is peaked exactly at $\langle Z_{12}\rangle=0+0i$ and is circularly symmetric in the $R+iX$ plane. Because the peak of the PDF is not necessarily at $\langle Z \rangle$, we generically call the most probable impedance $Z^\text{peak}$. In panels (b) and (d) of Fig.~\ref{Fig_PDF} we show the trajectories of $Z^\text{t}$ away from $Z^\text{peak}$ as the eight colored lines.

We also performed the same analysis on RMT numerical simulations generated using Eq.~\ref{EQN_ZRCM}. We create ensembles of $600$ realizations with different $\alpha$ values using eigenvalues $\zeta^\text{RMT}$ of Hamiltonians generated from a Gaussian Orthogonal Ensemble (GOE) that are of size $100000\times 100000$. Because RMT is a universal scattering theory, meaning it is insensitive to system specific details, agreement between our data and RMT predictions allow us to conclude that our results are not specific to just microwave cavities but can be applied to any complicated wave scattering systems such as those found in optics or acoustics.

\section{Discussion}\label{SEC_DISCUSSION}

\begin{figure}[htb]
\centerline{
\includegraphics[width=13cm]{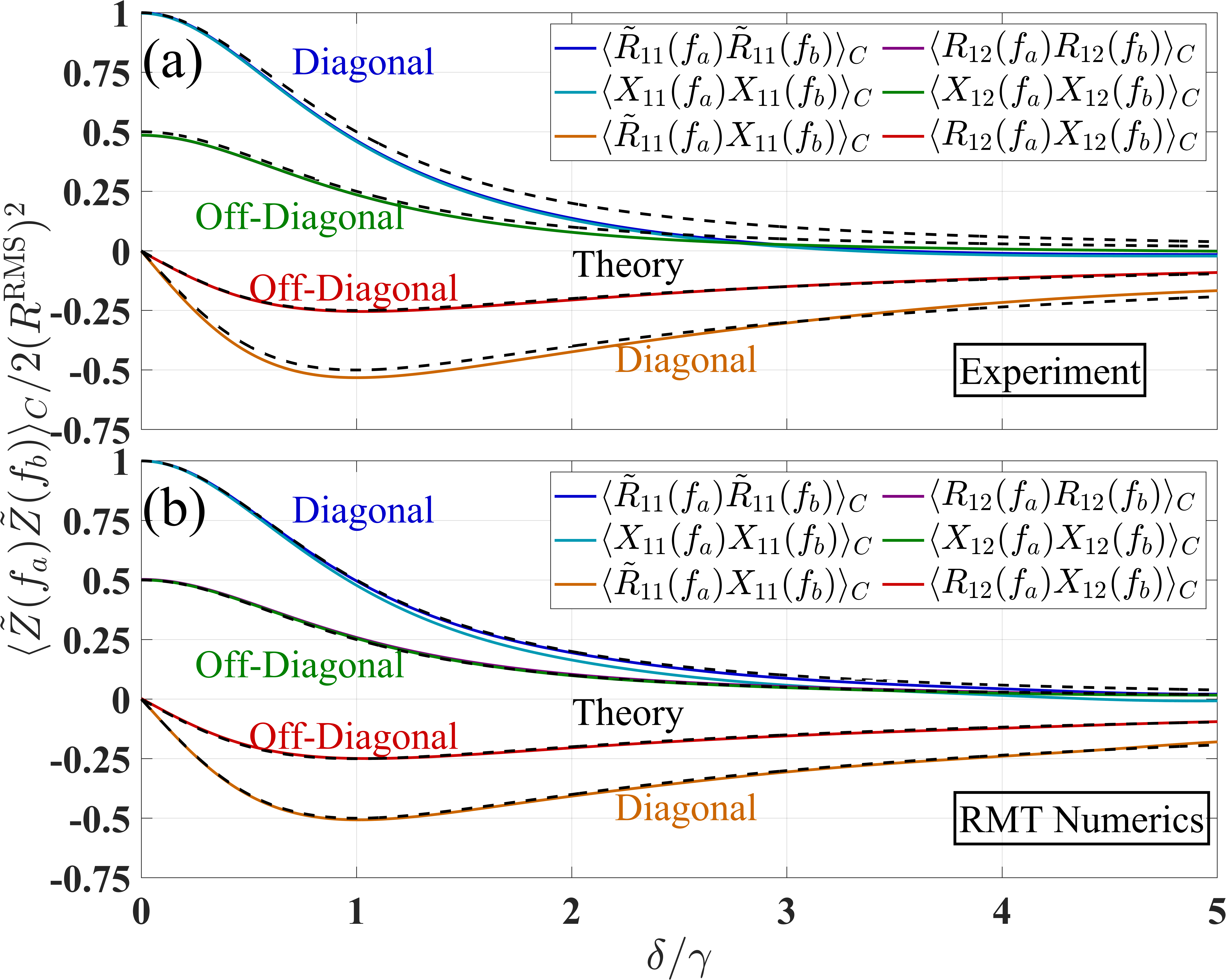}}
\caption{Two point frequency correlation functions of complex impedance normalized by twice $(R^\text{RMS})^2=\langle(R_{12})^2\rangle$, as a function of $\delta/\gamma=\frac{|f_a-f_b|}{2\alpha\Delta}$. Dashed black lines correspond to Eqs~\ref{EQN_CorrR}-\ref{EQN_CorrTCross} while solid colored lines are data results. (a) Experimental data from a three dimensional cavity with absorption $\alpha=5.5$, mean mode spacing $\Delta=0.06~\text{MHz}$ and characteristic frequency scale $\gamma=0.66~\text{MHz}$. (b) RMT numerical simulation with absorption $\alpha=5.5$, mean mode spacing $\Delta=14.2~\text{MHz}$ and characteristic frequency scale $\gamma=156.1~\text{MHz}$.}
\label{Fig_Corr}
\end{figure}

In this section we present the statistical results from the experiments and RMT numerics described in Sec.~\ref{SEC_EXP}, and compare these to the predictions derived in Sec.~\ref{SEC_THEORY}. Data from twenty six experimental and thirty one RMT numerical ensembles was prepared, but when appropriate only a characteristic example will be shown. Recall that all the results displayed are for the normalized universal impedance $Z$ calculated using Eq.~\ref{EQN_NORMZ}.

We first consider the two point frequency correlation functions, and show an example in Fig.~\ref{Fig_Corr}. The correlation functions have been normalized by $2(R^\text{RMS})^2=2\langle(R_{12})^2\rangle$.  In panel (a) we present experimental results from a three dimensional cavity with $\alpha=5.5$ while panel (b) has RMT numerical results from a simulation with the same loss. The correlation functions of the diagonal element, $\langle\tilde{R}_{11}(f_a)\tilde{R}_{11}(f_b)\rangle_C$, $\langle X_{11}(f_a)X_{11}(f_b)\rangle_C$, and $\langle\tilde{R}_{11}(f_a)X_{11}(f_b)\rangle_C$, are represented by the dark blue, light blue, and orange curves, respectively, in both panels. The correlation functions of the off-diagonal element, $\langle R_{12}(f_a)R_{12}(f_b)\rangle_C$, $\langle X_{12}(f_a)X_{12}(f_b)\rangle_C$, and $\langle R_{12}(f_a)X_{12}(f_b)\rangle_C$, are represented by the purple, green, and red curves. The dashed black lines are the expected correlation functions in the high loss limit using Eqs~\ref{EQN_CorrR}-\ref{EQN_CorrTCross}.

In both panels, we see the expected correlation at zero frequency separation (correlation of diagonal element is twice that of off-diagonal element) as well as the correct scaling of characteristic frequency $\gamma$ with increasing frequency separation $\delta$. As the frequency separation $\delta/\gamma$ increases, we see an increase in separation between the dashed black theory predictions and the data curves. This is to be expected as the number of data points available diminishes as $\delta$ increases, meaning we are more susceptible to fluctuations. However, over the entire range, the correlation functions of the off-diagonal element show better agreement with the theory. In deriving the equations presented in Sec.~\ref{SEC_THEORY}, we make the assumption that the PDF of the impedance is Gaussian distributed in the $R+iX$ plane and peaked at the mean value. This is always true for the off-diagonal impedance, but only true in the case of $\alpha\rightarrow\infty$ for the diagonal impedance. As can be seen in Fig.~\ref{Fig_PDF}(a-b), the diagonal impedance PDF is peaked away from the mean value of $1+0i$, and is not circularly symmetric. The lower the value of $\alpha$, the more the peak of the PDF moves towards $R_{pp}=0$. Despite the theory making the assumption that we are in the limit of high loss ($\alpha\gg1$), it still produces a good prediction for the two point frequency correlation functions for finite $\alpha$.

\begin{figure}[htb]
\centerline{
\includegraphics[width=15cm]{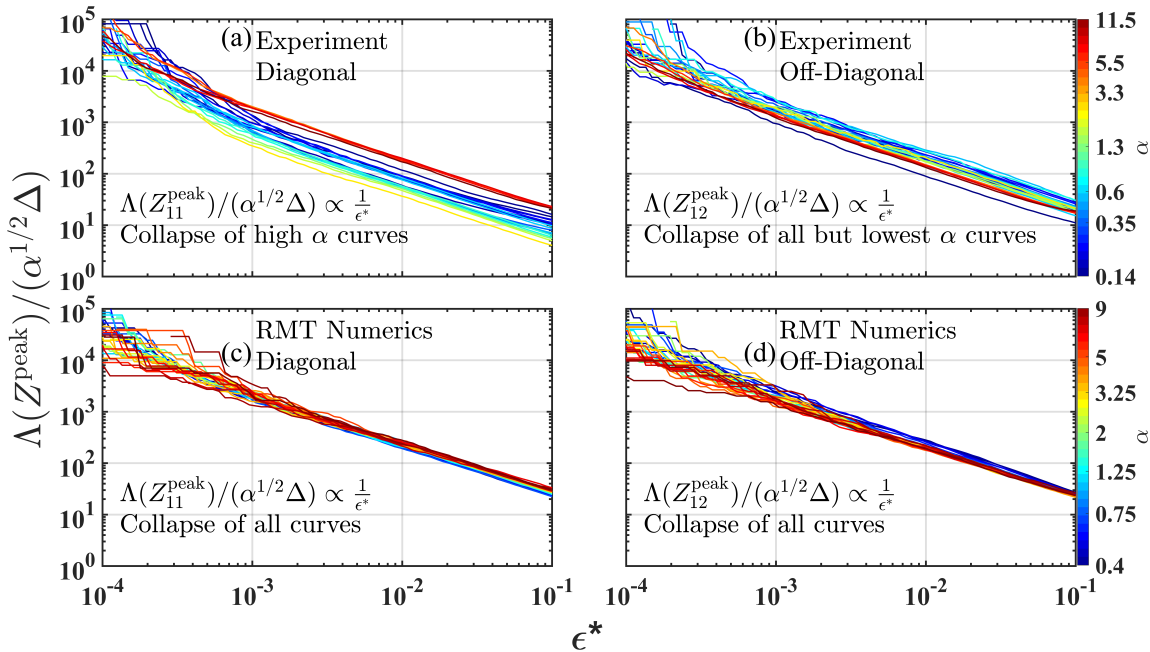}}
\caption{Average frequency interval $\Lambda$ between repeated instances of most probable impedance $Z^\text{peak}$ normalized by $\alpha^{1/2}\Delta$, as a function of tolerance $\epsilon$. Color scale corresponds to absorption parameter $\alpha$. (a-b) $\Lambda(Z^\text{peak})/(\alpha^{1/2}\Delta)$ from twenty six experimental ensembles, in diagonal and off-diagonal impedance cases, respectively. (c-d) $\Lambda(Z^\text{peak})/(\alpha^{1/2}\Delta)$ from thirty one RMT numerical ensembles, in diagonal and off-diagonal impedance cases, respectively.}
\label{Fig_Alphaepsilon}
\end{figure}

Next we investigate the question of the average frequency interval $\Lambda$ between repeated values of $Z^\text{t}$. Because $Z^\text{peak}$ is the most common impedance value, it stands to reason that $\Lambda(Z^\text{t})$ is minimized at $Z^\text{t}=Z^\text{peak}$. In Fig.~\ref{Fig_Alphaepsilon}, we plot $\Lambda(Z^\text{peak})/(\alpha^{1/2}\Delta)$ versus tolerance $\epsilon$ calculated from both experimental and RMT simulation data. If Eq.~\ref{EQN_FreqInterval_SIMP} is correct, all the curves should collapse and become degenerate to a single line described by $2(\pi^{1/2})/\epsilon$.

While we do indeed see a $\frac{1}{\epsilon}$ dependence for all the curves in each panel, there is some vertical offset between the curves from different ensembles, illustrated by the color scale corresponding to the ensemble $\alpha$ value. We have determined two possible reasons for the spread out nature of the curves. When looking at the experimental data in panels (a) and (b), it can be seen that the red curves corresponding to ensembles with $\alpha\geq5$ are degenerate, and only the blue, green and yellow curves fail to align with the scaling curve. This suggests that for small absorption there is some leftover $\alpha$ dependence to $\Lambda(Z^\text{peak})$ that is not properly captured by Eq.~\ref{EQN_FreqInterval_SIMP}, which assumes high loss ($\alpha\gg1$). However, the RMT numerical simulations in panels (c) and (d) result in $\Lambda(Z^\text{peak})/(\alpha^{1/2}\Delta)$ curves that seem degenerate for all $\alpha$ values. This implies the issue may not be in the theory but actually small inaccuracies in the $\Delta$ value used. Mode spacing in the RMT simulation is very well described by the prescribed $\Delta$, while for the experimental ensembles the value of $\Delta$ is just an approximation calculated using Weyl's formula. \rev{As for why the theory describes the statistical results of even low loss ensembles, both experimental and from RMT simulation, it is likely because the correlation functions remain qualitatively similar across all $\alpha$ values. Hence the simple geometrical argument for $\Lambda(Z^\text{t}=Z^\text{peak})=\frac{\pi\gamma R^\text{RMS}}{\epsilon}$ can be expected to hold.}

\begin{figure}[htb]
\centerline{
\includegraphics[width=15cm]{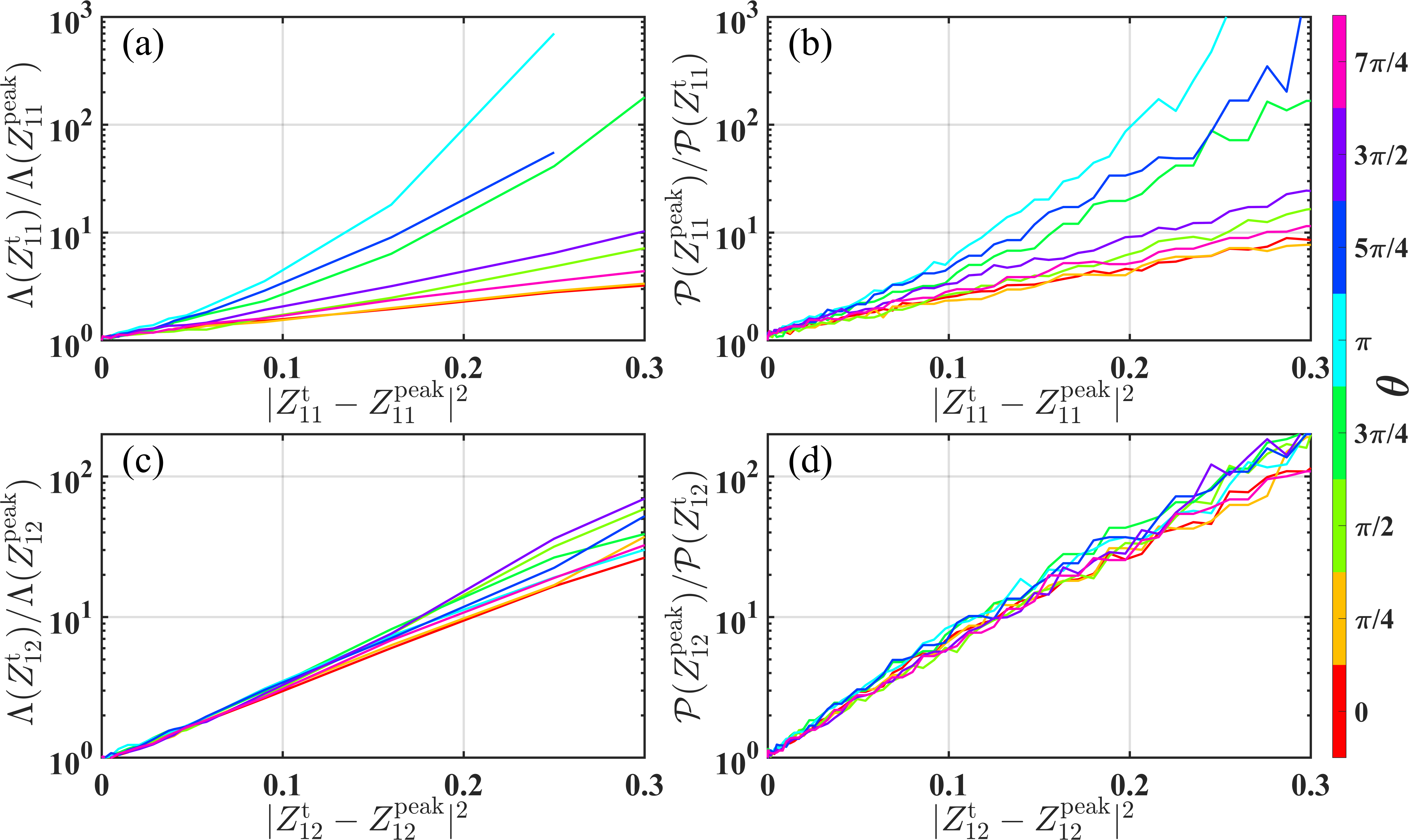}}
\caption{(a) and (c): Frequency interval $\Lambda(Z^\text{t})$ between repeated instances of $|Z-Z^\text{t}|\leq\epsilon=10^{-2}$, normalized by $\Lambda(Z^\text{peak})$ using the same tolerance. (b) and (d): Ratio $\mathcal{P}(Z^\text{peak})/\mathcal{P}(Z^\text{t})$. In all panels, the x-axis is the square difference between $Z^\text{t}$ and $Z^\text{peak}$ normalized by the radiation impedance. Color bar corresponds to the angles $\theta$ which are the same as the colored lines in Fig.~\ref{Fig_PDF}. Panels (a-b) show diagonal impedance $Z_{11}$, panels (c-d) show off-diagonal impedance $Z_{12}$. Data is from an experimental $\mathcal{D}=3,\alpha=5.5$ ensemble.}
\label{Fig_ZT}
\end{figure}

Having verified the relationship between $\Lambda(Z^\text{peak})$ and each of $\epsilon$,$\alpha$, and $\Delta$, we now look at the dependence of the repetition frequency $\Lambda$ on the target impedance $Z^\text{t}$. We plot $\Lambda(Z^\text{t})/\Lambda(Z^\text{peak})$ versus $|Z^\text{t}-Z^\text{peak}|^2$ in Fig.~\ref{Fig_ZT}(a) and (c), for diagonal and off-diagonal impedance, respectively. In this case, the allowed tolerance for counting a repetition is $\epsilon=10^{-2}$. The data comes from a three dimensional cavity with loss of $\alpha=5.5$, and the different colors correspond to the angle $\theta=\text{Arg}(Z^\text{t}-Z^\text{peak})$. Note that these are the same colors and angles as depicted in Fig.~\ref{Fig_PDF}(b) and (d) on top of the complex impedance PDFs.

In Fig.~\ref{Fig_ZT}(c), we see that $\Lambda(Z_{12}^\text{t})$ is independent of $\theta$, which is expected because of the circular symmetry of $\mathcal{P}(Z_{12})$. The eight curves in panel (c) appear linear on this log-lin scale, which is also expected because of the proportionality to the probability ratio $\mathcal{P}(Z_{12}^\text{peak})/\mathcal{P}(Z_{12}^\text{t})$. In Eq.~\ref{EQN_PDF}, we predicted that the ratio $\mathcal{P}(Z_{12}^\text{peak})/\mathcal{P}(Z_{12}^\text{t})$ would be an inverted bivariate Gaussian of the form $e^{(R_{12}^\text{t})^2+(X_{12}^\text{t})^2}$. As shown in panel (d), when plotted on a log-lin scale against the difference squared, the curves formed by this ratio are linear and look very similar to those in panel (c).

However, as was shown in Fig.~\ref{Fig_PDF}, the PDF of diagonal impedance $Z_{11}$ is not symmetric along the real axis. This means that both the distance from the peak $|Z^\text{t}-Z^\text{peak}|$ as well as the angle $\theta$ are important to the final value of $\Lambda(Z^\text{t}_{11})$. When the angle is within $\frac{\pi}{2}<\theta<\frac{3\pi}{2}$, the  $\Lambda(Z^\text{t}_{11})$ and $\mathcal{P}(Z^\text{peak}_{11})/\mathcal{P}(Z^\text{t}_{11})$ curves are concave up. In both panels (a) and (b), the light blue curve corresponding to $\theta=\pi$ abruptly stops at about $|Z^\text{t}-Z^\text{peak}|^2=0.25=0.5^2$. The reason is that for this ensemble with a loss of $\alpha=5.5$, there simply are virtually no occasions when the real part of the diagonal impedance is smaller than $R^\text{peak}_{11}-0.5\approx0.4$. 

Panels (a-b) and (c-d) of Fig.~\ref{Fig_ZT} show the direct relation between $\Lambda(Z^\text{t})$ and $\mathcal{P}(Z^\text{peak})/\mathcal{P}(Z^\text{t})$. Dividing the $\Lambda(Z^\text{t})/\Lambda(Z^\text{peak})$ curves in the left column by the corresponding $\mathcal{P}(Z^\text{peak})/\mathcal{P}(Z^\text{t})$ curves in the right column removes the multiple orders of magnitude change with $|Z^\text{t}-Z^\text{peak}|^2$. This means that when we use Eq.~\ref{EQN_FreqInterval_SIMP} to factor out the expected dependence of $\Lambda$ on the parameters $\alpha,~\Delta,~\epsilon$, and $Z^\text{t}$, we go from a frequency interval that varies over a range of $10^5$ (combining the y-scales of Figs.~\ref{Fig_Alphaepsilon}-\ref{Fig_ZT}) to a quantity that varies by only about a factor of 2. The fact we don't see a perfect equality between the two columns of Fig.~\ref{Fig_ZT} can again be explained by our theory being derived for the limiting case of high loss ($\alpha\gg1$), yet works suprisingly well even for moderate to low loss systems ($\alpha\approx1$).

\section{Conclusions} \label{SEC_CONCLUSION}
In this paper, we have presented an experimental study of the statistics of the RCM normalized impedance of microwave cavities. We have examined the two point frequency correlation functions, which agree nicely with the theoretical predictions. We also considered the question of how far on average one must look in frequency before seeing a repeated value of impedance. We demonstrated that this frequency interval $\Lambda$ depends on four parameters: the mean mode spacing $\Delta$, the absorption $\alpha$, the value of target impedance $Z^\text{t}$, and the tolerance $\epsilon$ on the desired impedance. Experimentally, \rev{the mode spacing} $\Delta$ and \rev{the mode overlap} $\alpha$ are inversely correlated and it can be difficult to tune one without affecting the other, which means it cannot be predicted whether $\Lambda$ will shrink or grow based on changes in just one of these parameters. Further, the PDF of $Z$ depends on $\alpha$, which introduces another, more complicated, dependence on loss to $\Lambda$. Using Random Matrix Theory simulations, we see statistical results that agree well with our data. The agreement between the experiment and RMT strongly imply our results are applicable to any sufficiently complicated wave scattering system, including those seen in optics or acoustics. An extension of this work is to consider not just special values of individual impedance elements, but special conditions on the entire $Z$ matrix, such as impedance matrix Exceptional Points, which are equivalent to scattering matrix Exceptional Points.

\section{\rev{Appendix A: Table of Parameters}} \label{SEC_APPTable}

%\rev{In this Appendix we provide a table with the parameters relevant to this work, including their definitions and relations to other parameters.}

\begin{table}[ht]
    \centering
    \begin{tabular}{|P{2cm}|P{6cm}|P{4cm}|}
    \hline
        Parameter Symbol & Definition & Relation to other Parameters\\
       \hline
         $Z^\text{raw}$ & Raw impedance matrix  & Eq.~\ref{EQN_S2Z} \\
         \hline
        $\langle Z^\text{raw} \rangle$ & Ensemble average of raw impedance. For a high quality ensemble with many realizations, becomes nearly identical to radiation impedance with small differences caused by short orbits. & N/A \\
        \hline
          $Z$ & Universal fluctuating impedance matrix, normalized to not contain system specific features & Eq.~\ref{EQN_NORMZ} \\
          \hline
          $Z^\text{RCM}$  & Universal fluctuating impedance matrix as predicted by the Random Coupling Model and RMT& Eq.~\ref{EQN_ZRCM} \\
          \hline
          $Z^\text{peak}$ & Impedance value that corresponds to the largest value of $\mathcal{P}(Z)$ & $\mathcal{P}(Z^\text{peak})=\text{max}[\mathcal{P}(Z)]$\\
          \hline
          $Z^\text{t}$ & Target impedance value of interest & N/A \\
          \hline
          $\epsilon$ & Tolerance on target impedance $Z^\text{t}$& N/A \\
          \hline
           \vspace{0.01cm} $R^\text{RMS}$ &\vspace{0.01cm} Root mean square fluctuation of the normalized impedance &\vspace{0.01cm} $R^\text{RMS}=\sqrt{\langle (R_{pp'})^2 \rangle}$ and at large loss $R^\text{RMS}=\sqrt{\frac{1}{\pi\alpha}}$\\
          \hline
           $\alpha$   & RCM loss parameter, quantifies degree of mode overlap & N/A \\
           \hline
          $\Delta$ & Mean frequency spacing of the modes & N/A \\
          \hline
          $\gamma$ & Mean spectral $Q$-width of the modes & $\gamma=2\alpha\Delta$\\
          \hline
         $\delta$ & Difference between two frequency values $f_{a}$ and $f_b$ & $\delta=|f_a-f_b|$ \\
         \hline
         $\Lambda$ & Average frequency interval between repeated occurances of target impedance value $Z^\text{t}$ & Eqs.~\ref{EQN_FreqInterval}-\ref{EQN_FreqInterval_SIMP} \\
         \hline
    \end{tabular}
    \caption{\rev{Table of important parameters used in this work. The symbol used is given in the first column and the definition in the second column. How that parameter is related to other parameters is given in the last column.}}
    \label{TABLE_PARA}
\end{table}

\clearpage
\newpage

\section{Appendix B: Random Coupling Model Loss Parameter Determination} \label{SEC_APPLOSS}

\begin{figure}[hbt]
\centerline{
\includegraphics[width=13cm]{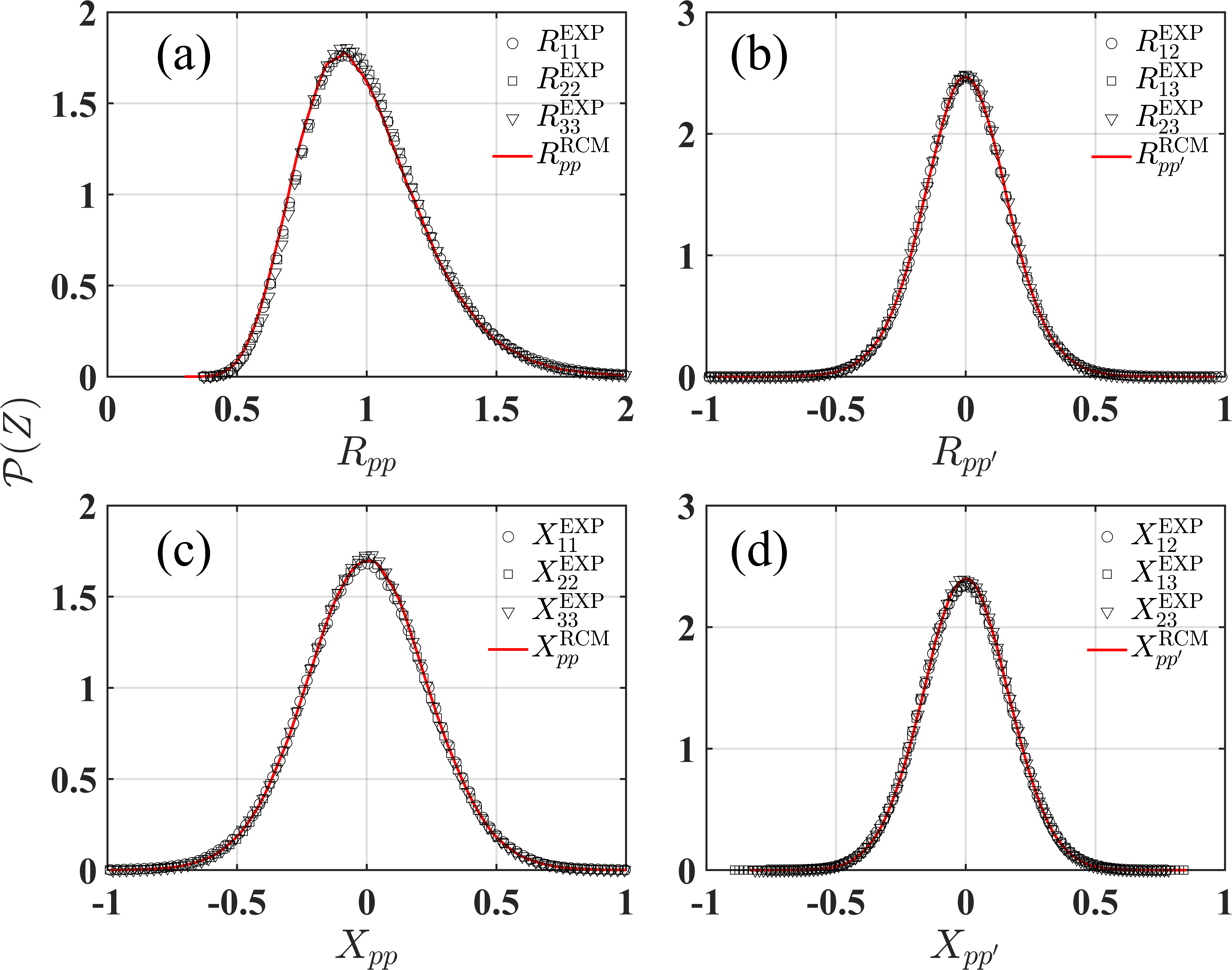}}
\caption{Black symbols correspond to the ensemble PDFs of the real and imaginary parts of elements of the RCM normalized impedance matrix from a three dimensional cavity. The red lines are the PDFs of $Z^{\text{RCM}}$ generated using Eq.~\ref{EQN_ZRCM} with loss parameter $\alpha=5.5$. All 12 unique (the system is reciprocal so $Z_{pp'}=Z_{p'p}$) experimental PDFs are simultaneously matched to the predicted PDFs, allowing for the determination of the cavity loss characteristic of the data.}
\label{Fig_ZAlpha}
\end{figure}

In this Appendix we demonstrate the validity of the $\alpha$ values assigned to the experimental ensembles used in this paper. We obtain a value for the loss parameter $\alpha$ of an ensemble by comparing PDFs of elements of the RCM normalized impedance matrix from Eq.~\ref{EQN_NORMZ} to the PDFs of numerically generated impedance using Eq.~\ref{EQN_ZRCM}. For a system with $M$ channels, there are $2M^2$ PDFs and we simultaneously match them all using a single $\alpha$ value. An example for the three dimensional cavity ensemble with $\alpha=5.5$ is shown in Fig.~\ref{Fig_ZAlpha}.

Because this process uses an ensemble that involves both cavity manipulation and a range of frequencies, we are in a sense averaging the contributions of multiple $\alpha$ values. This is reasonable when $\alpha$ is large and the contributions aren't too different from each other \cite{Fu2017}. However, since the PDFs of $Z$ have sensitivity of shape and variance to $\alpha$ that scale roughly with the inverse of $\alpha$, it can be difficult to do the same when $\alpha<1$.

Our experimental systems with their \textit{in situ} tunable perturbers allow us to make very high quality ensembles for which we can apply this process even in cases of very low loss. See the Supplemental Materials of Ref.~\cite{Shaibe2025_CTD} for an example where we simultaneously match all $8$ impedance PDFs of an $M=2$ channel graph to the RCM predicted PDFs with $\alpha=0.18$. This method is even applicable in the case of ultra low loss enabled through cryogenic environments, as shown in Ref.~\cite{Yeh2013Cyro}.

\section*{Acknowledgements}
We thank David Shrekenhamer and Timothy Sleasman of JHU APL for the design and fabrication of the metasurfaces used in this work. 
%This work was supported by ONR under grant/contract N000142312507. The views expressed are those of the author and do not reflect the official policy or position of the Department of Defense or the U.S. Government.  This work was also supported by NSF/RINGS under grant No. ECCS-2148318, and DARPA WARDEN under grant HR00112120021. 
This work was supported by NSF/RINGS under grant No. ECCS-2148318, ONR under grant N000142312507, and DARPA WARDEN under grant HR00112120021.

\bibliographystyle{unsrt}
\bibliography{ImpedanceStats}

\end{document}